\documentclass[prl,twocolumn,showpacs,floatfix]{revtex4}

\usepackage{graphicx}
\usepackage{amsmath}
\usepackage{amssymb}

\begin{document}
\title{Brownian refrigerator} 

\date{\today}

\author{C. Van den Broeck}
\affiliation{Hasselt University, B-3590 Diepenbeek,
Belgium}
\author{R. Kawai}
\affiliation{University of Alabama at Birmingham,
Birmingham, AL 35294, USA}

\begin{abstract}
Onsager symmetry implies that a Brownian motor, driven by a temperature
gradient, will also perform a refrigerator function upon loading. We
analytically calculate the corresponding heat flow for an exactly solvable
microscopic model and compare it with molecular dynamics simulations. 
\end{abstract}


\maketitle

Cooling techniques have a significant impact not only on our everyday life
but also on the advances in science. We have come a long way from the primitive
method of evaporative cooling, which our body conveniently uses when we
perspire,
over the evaporation by expansion of a cooling liquid in domestic refrigerators,
to high-tech methods including laser cooling, magnetic cooling, radiative
cooling or quantum cooling.  Temperature is a measure of thermal fluctuations
but the latter are not directly observable in macroscopic systems.  Recent
advances
in nanotechnology and molecular biology however allow to manipulate and even
manufacture
constructions on a molecular scale, where thermal fluctuations
can no longer be ignored.  To run such machines, we could copy the mode of
operation from
their macroscopic counterparts. An alternative, and arguably  more promising
approach, would be to utilize thermal fluctuations rather than fighting them. A
well documented example is the Brownian
motor~\cite{leibler94,reimann02},  which generates power through the
rectification of thermal fluctuations.
In this letter, we present a novel method of microscopic cooling based on a
Brownian motor in which, almost
paradoxically, thermal fluctuations themselves can be
harnessed to reduce the thermal jitter in one part of the system.

Our Brownian motor~\cite{vandenbroeck} (see Fig.~1)
consists of two parts, a triangle
and a flat paddle, which are rigidly linked and move as a single entity along
horizontal tracks. Its motion is induced, following Newton's laws, by the random
collisions with the
particles of the gas in which it is embedded.
Due to the random nature of these collisions, one expects a resulting rather
erratic motion of the motor.
The question of interest is whether, due to the asymmetry of the triangle, this
random motion is characterized by a nonzero average speed in a given
direction.
When both motor units reside in a single compartment with a gas (or liquid) at
equilibrium, we argue that such a sustained motion is impossible because it
would violate the second law of thermodynamics.  The construction would be a
{\it perpetuum mobile} of the second kind (or Maxwell demon), because the motion
could
be used, for example, to  lift a weight, hence extract work out of the single
heat bath.  This impossibility can more deeply be understood from the fact that
a system at equilibrium has a perfect  time-reversal symmetry.  Any movie of
such a system, or of any subpart of the system (like  monitoring  the
motion of our motor) is statistically indistinguishable when played forward or
backward in time.  Hence any form of systematic translation is
impossible.
\begin{figure}[b]
\includegraphics[width=8.6cm]{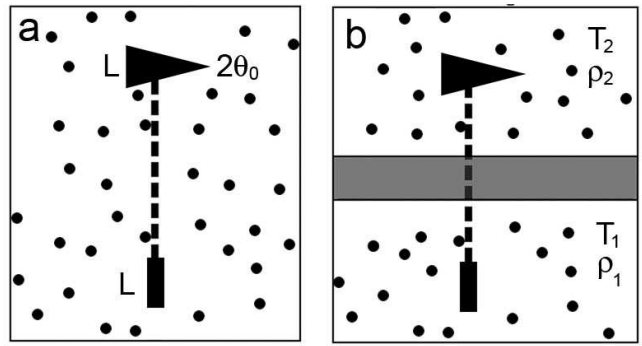}
\caption{An asymmetric object on horizontal tracks. (a) Performs Brownian
motion with no average systematic drift when in thermal
equilibrium. (b) Sustained average systematic motion when subject to a thermal
gradient.}
\label{fig:motor}
\end{figure}

Having concluded that no sustained motion occurs at equilibrium, we modify the
construction by inserting a perfectly insulating wall between both motor units
while not hampering their joint motion as illustrated in Fig.~1\textbf{b}. A
nonequilibrium state is now easily
realized by assuming that the particles in the separate lower and upper
compartments are at a {\it different} temperature, respectively $T_1$ and
$T_2$.  To intuitively understand what happens in this case, we
first
consider the extreme case $T_1>T_2=0$, corresponding to fully stationary
particles
(no thermal motion) in  reservoir 2 (upper reservoir with the triangle unit). 
When a particle hits the
paddle in reservoir 1, it transfers horizontal momentum to the motor.
It will do so in random quantities and equally so from the left as from the
right. This momentum will be dissipated as the triangle hits the stationary
particles in reservoir 2.  However,  it is clear that the resulting slowdown
will be much stronger when the motion of the motor is towards the left, with the
flat
face of the triangle transferring large momentum to the stationary particles. 
When the motion is towards the right, the inclined sides of the triangle
transfer less
momentum allowing it to move farther, in the same way as a sharp arrow can
penetrate deeper. We conclude that an average systematic motion is expected
pointing to the right in the direction of the sharp angle.  

The above handwaving arguments are confirmed by an analytic calculation,
which is
furthermore exact in the limit of
dilute gases~\cite{vandenbroeck}.
The average velocity of the motor is found to be  : 
\begin{eqnarray}
\overline{V}  &=&\rho_1 \rho_2
(1-\sin^2 \theta_0) \sqrt{\frac{m}{M}} \sqrt{\frac{\pi
k_B}{2M}} \nonumber \\
&&\times \frac{(T_1 - T_2) \sqrt{T_1}}{[2\rho_1 \sqrt {T_1} +
\rho_2\sqrt{T_2} (1 + \sin \theta_0)]^2}. 
\label{eq:V2}
\end{eqnarray}
Here $m$ and $M$ are the mass of gas particles and motor respectively
and $M \gg m$ is assumed.  The densities of gas
particles in reservoirs 1 and 2, $\rho_1$ and $\rho_2$, are assumed
to be small.  The shape of the motor units is defined by the
vertical cross section $L$ 
and the half apex angle of the triangle $\theta_0$ (see Fig.~1\textbf{a}).  In
agreement
with the
above discussion, we find for  $T_1>T_2$ a positive average velocity, i.e., a
velocity
to the right. At equilibrium, $T_1=T_2$, the velocity is zero (and also  for 
$\theta_0=\pi/2$, in agreement with the fact that the triangle reduces to a
flat
paddle without preferred direction). Note furthermore that equilibrium is a
point of flow reversal: for  $T_1<T_2$, the motor has a negative velocity.
Somewhat surprisingly, it then moves towards the left, i.e., in  a direction
opposite to the sharp angle!

The transition from a Brownian motor to a Brownian refrigerator can be made by
invoking a general stability principle that appears in several branches of
physics under different names: Newton's action-reaction law in mechanics,
Lenz's law of magneto-induction in electromagnetism and, relevant to the problem
under
consideration here, the Le Chatelier-Braun principle in thermodynamics.  The
latter states that an action on a system at equilibrium induces processes that
attenuate or counteract the original perturbation.  This principle of negative
feedback is clearly an expression of stability of the original equilibrium
state.
Now, consider again the above construction, but at equilibrium with $T_1=T_2$.
We perturb this state by applying an external force $F<0$  on the motor, moving
it to the left.  According to the above principle, we expect that the system
reacts by a process that induces an opposing motion to the right. In the above
discussion we have identified such a process: we need to increase the
temperature in reservoir 1 with respect to reservoir 2. This
entails the appearance of a heat flow from reservoir 2 to 
reservoir 1, cooling down reservoir 2. Similarly, an opposite heat flow from
reservoir 1 to reservoir 2 will arise when applying a force $F>0$.

The above discussion may appear wishful and does not reveal neither
the mechanism nor the amplitude of the process by which this miraculous heat
flow appears.  Fortunately, a simple but basic argument from the theory of
linear irreversible thermodynamics~\cite{prigogine} confirms the
prediction and allows us to evaluate  the strength of the phenomenon. 
We note that the Brownian motor, described above, is the result of a
cross-effect: a thermal gradient $\Delta T = T_1-T_2$  produces an average motor
velocity  $\overline{V}$.  In irreversible thermodynamics, one calls $J_1 =
\overline{V}$ a
flow and $X_2 = 1/T_2-1/T_1$ a thermodynamic force~\cite{callen60}. Since the
flow is zero at
equilibrium, i.e., in the absence of the thermodynamic force, we expect  to
find, for a small temperature
difference $\Delta T$, $T_1=T+\Delta T/2$, $T_2=T-\Delta T/2$,  a linear
relation  between flow $J_1$ and force $X_2 = \Delta T/T^2$, namely:  $J_1 =
L_{12} X_2$.  The value of the coefficient $L_{12}$ is found from Eq
(\ref{eq:V2}):
\begin{eqnarray}
L_{12} &=& \rho_1 \rho_2
(1-\sin^2 \theta_0) \sqrt{\frac{m}{M}} \sqrt{\frac{\pi
k_B}{2M}} \nonumber\\
&&\times\frac{T^{3/2}}{[2\rho_1 +
\rho_2 (1 + \sin \theta_0)]^2}.
\label{eq:L12}
\end{eqnarray}

Cross-processes are no exception.  Well-documented cases are the 
Seebeck, Thomson and Peltier effects, where temperature differences induce
electrical currents or vice-versa. The important observation in our discussion
is that Onsager~\cite{onsager31} has identified a profound symmetry relation
between any cross-process and its mirror process, deriving from the time
reversal symmetry of underlying microscopic dynamics. More precisely, for any
relation $J_1 = L_{12} X_2$, there is a mirror relation $J_2 = L_{21}  X_1 $
with an identical proportionality coefficient  $L_{21} = L_{12}$.
In our case,  $J_2$ is the flow associated to the temperature gradient $X_2$,
i.e.,  it is a heat flow $\dot{Q}_{1 \rightarrow 2}$  from reservoir 1
to reservoir 2. 
The particle flow is associated with a difference in the chemical potential. In
our setting, the latter is produced by the application of an
external mechanical force $F$.  The corresponding thermodynamic force is
$X_1=F/T$.
The relation $J_2 = L_{21} X_1$,  with  Eq.~(\ref{eq:L12}),
implies that the heat flow $J_2=\dot{Q}_{1 \rightarrow 2}$  is given by:
\begin{eqnarray}
\dot{Q}_{1 \rightarrow 2}&=&  \rho_1 \rho_2
(1-\sin^2 \theta_0)\, \sqrt{\frac{m}{M}}\, \sqrt{\frac{\pi
k_B T}{2M}} \nonumber\\
&&\times \frac{F}{[2\rho_1 +
\rho_2 (1 + \sin \theta_0)]^2}.
\label{eq:Q}
\end{eqnarray}
This heat flow is quite large, namely of the order of the power delivered by
the external force (thermal speed of the motor, $\sqrt{k_BT/M}$, times $F$)
multipled by
$\sqrt{m/M}$.  Furthermore, 
in agreement with our foregoing discussion based on stability principles, the
direction of heat transfer depends on the direction of the force $F$, in such a
way that it activates an opposing Brownian motor. 

While heat is transferred from one reservoir to another, we have to realize
that the motion induced by the
force will have another, quite familiar effect, namely dissipation by the
frictional force $\gamma \overline{V}$, where $\gamma$ is a frictional
coefficient.
The dissipation will heat up both reservoirs and thus it may forestall the
cooling. This "Joule" energy stems from the
power input $P = F \overline{V}$. But with the velocity
itself proportional to the applied force $\overline{V} = 
F/\gamma$, the dissipation is proportional to $F^2$. The quadratic law conforms
with the fact that the dissipation is
independent of the direction of the applied force. We conclude that the cooling
effect is dominant for small forces.
The application of such forces along with the use of either small motor
molecules ($M$ small) or larger substrate particles ($m$ large) will induce a
significant
cooling effect in one of the compartments.

All the above results, the explicit value of the Onsager coefficients and in 
particular Onsager symmetry, are confirmed by a direct microscopic calculation. 
The starting point for this derivation is the following Boltzmann-Master 
equation for the probability distribution $P(V,t)$ for the speed of the motor:
\begin{eqnarray}
&&\partial_t P(V,t) =-\frac{F}{M}\frac{\partial}{\partial V} P(V,t) \nonumber\\ 
&&+\int dV' [ W (V|V') P(V',t)-W(V'|V) P(V,t)]
\label{eq:Boltzmann}
\end{eqnarray}
where $W(V|V')$ is a transition probability per unit time for the motor to 
change velocity from $V'$ to $V$ due to the collisions with the gas particles.  
This approach is exact in the limit of dilute gases. Analytic expressions of 
the transition probability for general convex objects are availabe for the 
dilute gases~\cite{vandenbroeck}.
 
Expanding Eq.~(\ref{eq:Boltzmann}) in Taylor series up to second order in 
$\sqrt{m/M}$, we find the following explicit result (to linear order in $F$ and 
$\Delta T$):
\begin{eqnarray}
J_1 &=& L_{11} X_1 + L_{12} X_2 \nonumber \\ J_2 &=& L_{21} X_1 + L_{22} X_2
\label{eq:JLX}
\end{eqnarray}
with $J_1$, $J_2$, $X_1$ and $X_2$ as defined before, and
\begin{equation}
L_{11} = \frac{T}{\gamma}, \quad L_{22} = \frac{k_B \gamma_1 
\gamma_2 T^2}{M\gamma} .
\label{eq:L}
\end{equation}
The friction coefficients in each of the reservoir are given by
\begin{equation}
\gamma_1 = 8\rho_1 L \sqrt{\frac{k_B T m}{2\pi}}, \quad \gamma_2 = 4\rho_2 L 
\sqrt{\frac{k_B T m}{2\pi}} (1 + \sin\theta_0) 
\end{equation}
and the total friction by $\gamma = \gamma_1+\gamma_2$.
$L_{12}$ is as given in Eq.~(\ref{eq:L12}) and Onsager symmetry, 
$L_{12}=L_{21}$, is confirmed.

The above result also supplies us with the explicit expression for Joule
heating in each reservoir:
\begin{equation}
\dot{Q}_{Ji} = \gamma_i \overline{V}^2 ={\gamma_i F^2}/{\gamma^2}, 
\quad (i=1,2).
\end{equation}

The ratio of the heat transferred
from reservoir 2 to 
reservoir 1 ($\dot{Q}_{2 \rightarrow 1}= -\dot{Q}_{1 \rightarrow 2}$) over  the
Joule heat
dissipated in reservoir 2 is thus found to be
\begin{equation}
\frac{\dot{Q}_{2 \rightarrow 1}}{\dot{Q}_{J2}}= -2(1-\sin\theta_0)\, k_B T\, 
\frac{m}{M}\,
\frac{L\rho_1}{F} .
\label{eq:ratio}
\end{equation}
If this ratio is larger than one, reservoir 2 will cool down. The following
condition ensues in terms of the applied small (negative) force:
\begin{equation}
0 > F > -2 (1-\sin\theta_0)\, k_B T\,  \frac{m}{M}\,  L\rho_1 .
\label{eq:condition}
\end{equation}
A similar condition applies for the range of small positive forces, inducing
cooling in reservoir 1. 

Besides the issue of Joule heating, we need
also address the fact that the Brownian motor itself conducts
heat~\cite{conductivity}. From Eqs. (\ref{eq:JLX}) and (\ref{eq:L}) we
obtain the Fourier law, $L_{22} X_2 = \kappa
\Delta T$ with $\kappa = L_{22} / T^2$.
Assuming there is no other conductivity leak involved, one concludes from
Eqs.~(\ref{eq:L12}), (\ref{eq:JLX}) and (\ref{eq:L}), that total heat flow
($J_2=0$) will vanish at a relative
temperature gradient $\Delta T/T$ of the order of the ratio of the
drift speed $F/\gamma$ of the motor over the thermal speed $\sqrt{k_B T/m}$ of
the gas particles. Note that this result in independent of the mass $M$ of the
motor due to the fact that both heat currents, originating from thermal
fluctuations, share the same $1/{M}$ dependence.

\begin{figure}[b]
\includegraphics[width=8.6cm]{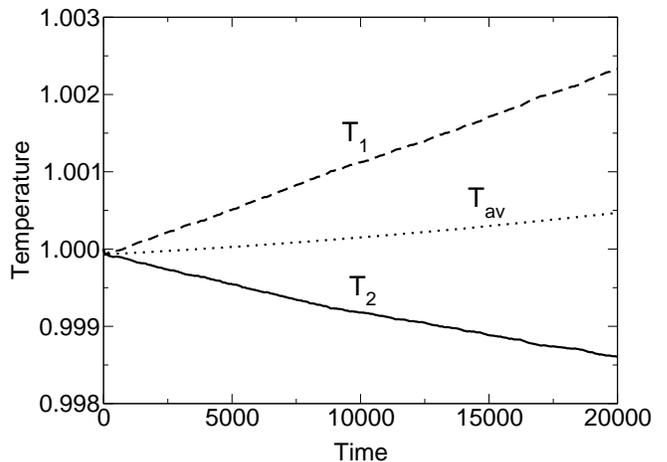}
\caption{The motor of mass $M=2m$ is driven by a constant force satisfying Eq.
(\ref{eq:condition}), namely $F=-0.007$. 
The temperature of two reservoirs, $T_1$ (dashed line) and $T_2$ (solid line)
are initially equal to $1$.   Clearly the
temperature of reservoir 2 ($T_2$) is decreasing, while reservoir 1 is heating
up. Note
that the average temperature $T_{av} = (T_1+T_2)/2$ increases due
to the Joule heat
dissipated in both reservoirs.}
\label{fig:T}
\end{figure}

To test the above predictions in a more realistic model, we have performed
extensive two-dimensional hard disk molecular dynamics simulations (diameter
$\sigma=1$). The system comprizes two separate rectangles
(size=$1200 \times 300$), with  periodic boundary conditions in both
directions, each containing 3600 hard disks (density
$\rho_1 = \rho_2 =0.01$).  The motor unit is as represented in 
Fig.~1 ($L=10$ and $\theta_0=10^\circ$).  Averages are
taken over
30000 realizations.
In Fig.~2, we show the evolution of the
temperatures in both reservoirs, starting with $T_1=T_2=1$ (dimensionless
units $k_B=m=1$ and $M=2$ are used). We apply
a weak force $F=-0.007$ to the motor.  The cooling effect in 
reservoir 2 is clearly
observed.  To compare
the theoretical results with the simulations, we plot in Fig.~3 the net heat
flow
injected into reservoir 2, $\dot{Q}_{1 \rightarrow 2}+\dot{Q}_{J2}$.
The agreement between theory and simulation is, as expected, very good for
large values of $M/m$, but even for small values of $M/m$ the order of magnitude
 is correctly predicted
by the theory.

\begin{figure}[t]
\includegraphics[width=8.6cm]{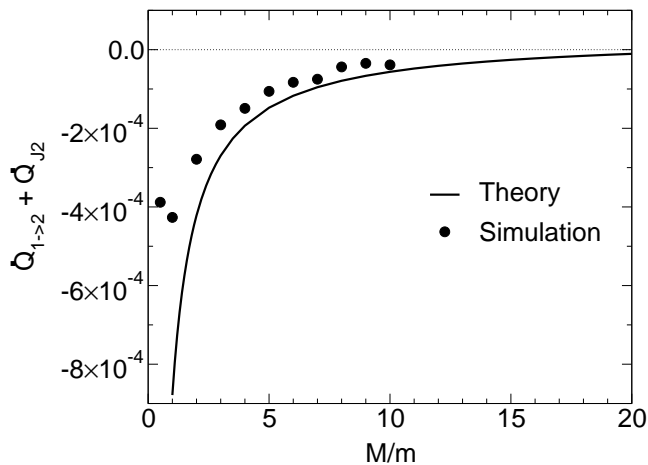}
\caption{The rate of energy change in reservoir 2 as a function of the
 mass of the motor  upon application of a force $F=-0.007$.  The energy change
encompasses  both the heat transfer from
reservoir 2 into 
reservoir 1 as well as the  Joule heating in reservoir 2.  Cooling is clearly
observed.}
\label{fig:flux}
\end{figure}
\begin{figure}[b]
\includegraphics[width=8.6cm]{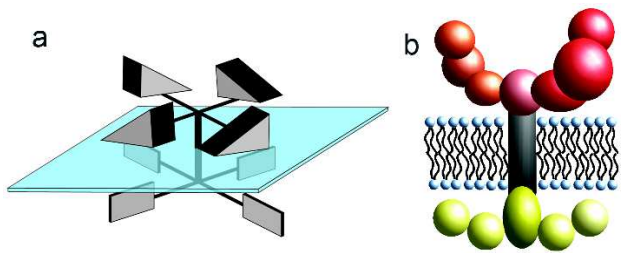}
\caption{A rotational version of the proposed
refrigerator. (a) A chiral
rotor is placed across an insulating wall. (b) A chiral biological
molecule will channel heat across a membrane when a small external torque is
applied.}
\end{figure}

Moving beyond the academic discussion, we need to address a number of issues
of more experimental or technological relevance.
Under which conditions could the above discussed heat flow be observed or
applied? The
explicit result given in Eq.~(\ref{eq:Q}) was
derived for an ideal gas. Molecular dynamics carried out at larger densities
however show
that the order of magnitude of the heat flow remains the same. We can
therefore use the above result to estimate the effect, even when we have in mind
motors operating in an aqueous or lipid environment. Consider a
macromolecule embedded in lipid bilayers, under physiological
conditions, $T \approx 300$ K. For a typical
value of its mass $M=10^5m$ ($m$
being typically the mass of a water molecule) and for a force of 
0.1 pN, Eq.~(\ref{eq:Q}) predicts ($\rho_1=\rho_2$) a  heat
flow of the order of $5 \times 10^{-17}$ J/sec.   On the other hand, using a
typical value for the drag coefficient of
a protein
$\gamma \sim 100 \text{pN} \cdot \text{s/m}$, the 
Joule heat is found to be of the same order, and both
heat effects are comparable. Hence, a negative net heat
flow of this order of magnitude 
will appear for $F$ slightly smaller or $M$ slightly  larger. Such a flow will
cool down an aqueous reservoir of diameter $0.1 \mu$m by one degree in about one
minute time.

We finally note that an alternative construction with rotational instead of
translational motion of the motor would appear easier to realize. One can
imagine a chiral molecule protruding on both sides of a membrane that is set
into rotational motion by some polarizing force, see Fig.~4. As we have
revealed above,  such a construction could in principle function as a Brownian
refrigerator, cooling down one side and heating up the other. Putting such
constructions in parallel would amplify the heat current. Putting them
in series would allow to channel heat out of the system following
a predesigned pathway. Aside from the possible application in 
molecular biology, the aforementioned advances in nanotechnology should allow to
construct
carefully designed devices of the type described above. As a final comment, we
stress that the Le Chatelier-Braun principle implies the appearance of a
Brownian refrigerator under very general conditions, whenever an opposing
thermal Brownian motor can be activated in response to an applied external or
thermodynamic force.

We thank Chris Jarzynski for helpfull discussions.

\
\end{document}